\documentclass[fleqn,usenatbib]{mnras}

\usepackage{newtxtext,newtxmath}
\usepackage[T1]{fontenc}
\usepackage{ae,aecompl}

\usepackage{graphicx}	% Including figure files
\usepackage{amsmath}	% Advanced maths commands
\usepackage{amssymb}	% Extra maths symbols
\usepackage{calc}
\usepackage{ifthen}

\newcommand{\idealizedHaloScatterFit}{
\begin{eqnarray}
 \log_{10} \sigma_{\log_{10}c_\mathrm{200c}} &=& a(c_\mathrm{200c}) + b \log_{10} N_\mathrm{h} \nonumber \\
 a(c_\mathrm{200c}) &=& -1.04 +2.03 \log_{10} c_\mathrm{200c}  -0.53 \log_{10}^2 c_\mathrm{200c} \nonumber \\
 b    &=& -0.47 \pm 0.06,
 \label{eq:idealizedHaloScatterFit}
\end{eqnarray}
}

\newcommand{\cosmologicalHaloScatterFit}{
\begin{eqnarray}
 \log_{10} \sigma_{\log_{10}c_\mathrm{200c}} &=& a(c_\mathrm{200c}) + b \log_{10} N_\mathrm{h} \nonumber \\
 a(c_\mathrm{200c}) &=& -0.20 +1.46 \log_{10} c_\mathrm{200c}  -0.25 \log_{10}^2 c_\mathrm{200c} \nonumber \\
 b    &=& -0.54 \pm 0.06.
 \label{eq:cosmologicalHaloScatterFit}
\end{eqnarray}
}

\newcounter{MDPLDone}
\def\MDPL{\ifthenelse{\equal{\arabic{MDPLDone}}{0}}{MultiDark Planck N-body (MDPL2)\setcounter{MDPLDone}{1}}{MDPL2}}

\newcounter{MCMCDone}
\def\MCMC{\ifthenelse{\equal{\arabic{MCMCDone}}{0}}{Markov Chain Monte Carlo (MCMC)\setcounter{MCMCDone}{1}}{MCMC}}

\title[Halo concentrations from extended Press-Schechter]{Halo concentrations from extended Press-Schechter merger histories}

\author[A. J. Benson et al.]{
Andrew J. Benson,$^{1}$\thanks{E-mail: abenson@carnegiescience.edu}
Aaron Ludlow,$^{2}$
and Shaun Cole$^{3}$
\\
$^{1}$ Carnegie Observatories, 813 Santa Barbara Street, Pasadena, CA 91101, USA\\
$^{2}$ International Centre for Radio Astronomy Research, 7 Fairway, Crawley, 6009, Perth, WA, Australia\\
$^{3}$ Institute for Computational Cosmology, Department of Physics, Science Laboratories, Durham University, South Road, Durham,\\ DH1 3LE, UK
}

\date{Accepted XXX. Received YYY; in original form ZZZ}

\pubyear{2018}

\begin{document}
\newcommand{\ludlowFourteenNoEnvironmentConcentrationMean}{1.077} % mean of log-concentration
\newcommand{\ludlowFourteenNoEnvironmentConcentrationScatter}{0.133} % scatter of log-concentration
\newcommand{\ludlowFourteenNoEnvironmentConcentrationMeanTable}{1&&077} % mean of log-concentration
\newcommand{\ludlowFourteenNoEnvironmentConcentrationScatterTable}{0&&133} % scatter of log-concentration
\newcommand{\ludlowFourteenNoEnvironmentExponentCollapseThreshold}{+0.000} % additional exponent of growth function in collapse threshold
\newcommand{\ludlowFourteenEnvironmentConcentrationMean}{1.051} % mean of log-concentration
\newcommand{\ludlowFourteenEnvironmentConcentrationScatter}{0.140} % scatter of log-concentration
\newcommand{\ludlowFourteenEnvironmentConcentrationMeanTable}{1&&051} % mean of log-concentration
\newcommand{\ludlowFourteenEnvironmentConcentrationScatterTable}{0&&140} % scatter of log-concentration
\newcommand{\ludlowFourteenEnvironmentExponentCollapseThreshold}{+0.071} % additional exponent of growth function in collapse threshold
\newcommand{\ludlowFourteenNBodyConcentrationMean}{1.084} % mean of log-concentration
\newcommand{\ludlowFourteenNBodyConcentrationScatter}{0.152} % scatter of log-concentration
\newcommand{\ludlowFourteenNBodyConcentrationMeanTable}{1&&084} % mean of log-concentration
\newcommand{\ludlowFourteenNBodyConcentrationScatterTable}{0&&152} % scatter of log-concentration
\newcommand{\ludlowFourteenNBodyMassMinimum}{ 9.402} % minimum log10 M200 mass of halos included in sample
\newcommand{\ludlowFourteenNBodyMassMaximum}{ 9.902} % maximum log10 M200 mass of halos included in sample

\newcommand{\ludlowSixteenNoEnvironmentConcentrationMean}{1.104} % mean of log-concentration
\newcommand{\ludlowSixteenNoEnvironmentConcentrationScatter}{0.103} % scatter of log-concentration
\newcommand{\ludlowSixteenNoEnvironmentConcentrationMeanTable}{1&&104} % mean of log-concentration
\newcommand{\ludlowSixteenNoEnvironmentConcentrationScatterTable}{0&&103} % scatter of log-concentration
\newcommand{\ludlowSixteenNoEnvironmentExponentCollapseThreshold}{+0.000} % additional exponent of growth function in collapse threshold
\newcommand{\ludlowSixteenEnvironmentConcentrationMean}{1.077} % mean of log-concentration
\newcommand{\ludlowSixteenEnvironmentConcentrationScatter}{0.114} % scatter of log-concentration
\newcommand{\ludlowSixteenEnvironmentConcentrationMeanTable}{1&&077} % mean of log-concentration
\newcommand{\ludlowSixteenEnvironmentConcentrationScatterTable}{0&&114} % scatter of log-concentration
\newcommand{\ludlowSixteenEnvironmentExponentCollapseThreshold}{+0.071} % additional exponent of growth function in collapse threshold
\newcommand{\ludlowSixteenNBodyConcentrationMean}{1.084} % mean of log-concentration
\newcommand{\ludlowSixteenNBodyConcentrationScatter}{0.152} % scatter of log-concentration
\newcommand{\ludlowSixteenNBodyConcentrationMeanTable}{1&&084} % mean of log-concentration
\newcommand{\ludlowSixteenNBodyConcentrationScatterTable}{0&&152} % scatter of log-concentration
\newcommand{\ludlowSixteenNBodyMassMinimum}{ 9.402} % minimum log10 M200 mass of halos included in sample
\newcommand{\ludlowSixteenNBodyMassMaximum}{ 9.902} % maximum log10 M200 mass of halos included in sample

\newcommand{\parameterBestFitf}{$ +0.061^{+  0.020}_{-  0.005}$} % best-fit value for parameter f
\newcommand{\parameterBestFitp}{$ +0.218^{+  0.045}_{-  0.130}$} % best-fit value for parameter p
\newcommand{\parameterBestFitalpha}{$ +0.077^{+  0.017}_{-  0.009}$} % best-fit value for parameter alpha
\newcommand{\parameterBestFitC}{$ +625.0^{+   77.0}_{-   15.0}$} % best-fit value for parameter C
\newcommand{\parameterBestFita}{$ +0.791^{+  0.074}_{-  0.023}$} % best-fit value for parameter a
\newcommand{\parameterBestFitGZero}{$ +0.591^{+  0.009}_{-  0.010}$} % best-fit value for parameter GZero
\newcommand{\parameterBestFitgammaOne}{$ +0.253^{+  0.019}_{-  0.019}$} % best-fit value for parameter gammaOne
\newcommand{\parameterBestFitgammaTwo}{$ +0.124^{+  0.017}_{-  0.019}$} % best-fit value for parameter gammaTwo
\newcommand{\parameterBestFitA}{$ +0.302^{+  0.013}_{-  0.003}$} % best-fit value for parameter A

\label{firstpage}
\pagerange{\pageref{firstpage}--\pageref{lastpage}}
\maketitle

\begin{abstract}
We apply the model relating halo concentration to formation history proposed by Ludlow et al. to merger trees generated using an algorithm based on excursion set theory. We find that while the model correctly predicts the median relation between halo concentration and mass, it underpredicts the scatter in concentration at fixed mass. Since the same model applied to N-body merger trees predicts the correct scatter, we postulate that the missing scatter is due to the lack of any environmental dependence in merger trees derived from excursion set theory. We show that a simple modification to the merger tree construction algorithm, which makes merger rates dependent on environment, can increase the scatter by the required amount, and simultaneously provide a qualitatively correct correlation between environment and formation epoch in the excursion set merger trees.
\end{abstract}


%%%%%%%%%%%%%%%%%%%%%%%%%%%%%%%%%%%%%%%%%%%%%%%%%%

%%%%%%%%%%%%%%%%% BODY OF PAPER %%%%%%%%%%%%%%%%%%

\section{Introduction}

It is now very well established that the density profiles of cold dark matter (CDM) halos can be well described by simple, universal forms such as the NFW profile \cite{navarro_structure_1996,navarro_universal_1997} or, somewhat more accurately, by the Einasto profile \citep{gao_redshift_2008,ludlow_einasto_2017}. The density profile is thought to play a significant role in shaping the properties of galaxies which form in dark matter halos, as it contributes significantly to the rotation curve, and gravitational potential.

In semi-analytic models of galaxy formation, the density profile of the halo directly affects determinations of galaxy sizes \citep{cole_hierarchical_2000,jiang_is_2018}, and in many models affects the mass loading of outflows from galaxies \citep{cole_hierarchical_2000,benson_galacticus:_2012}. While semi-analytic models of galaxy formation are most often applied to dark matter halo merger trees extracted from N-body simulations, from which dark matter profiles can be measured directly, they can also be applied to merger trees generated using other techniques, such as those based on excursion set theory \citep{bond_excursion_1991,lacey_merger_1993,cole_hierarchical_2000,parkinson_generating_2008}. Such approaches have some advantages over the use of N-body merger trees, for example allowing much finer time resolution to be achieved (which can affect the results of semi-analytic models; \citealt{benson_convergence_2012}), rapid exploration of different dark matter physics \citep{benson_dark_2013}, and avoidance of numerical noise issues which occur in N-body halos at low particle number \citep{benson_mass_2017}.

However, merger trees built from excursion set theory have, so far, only provided halo masses---they say nothing about other key physical properties of halos such as their density profiles, spins, or environment. Typically, these quantities are assigned to halos in such trees by appealing to an empirical correlation with halo mass and redshift, or by drawing at random from some measured distribution. In the case of halo concentrations for example, the usual approach is to assign a concentration based on a concentration-mass-redshift relation measured from an N-body simulation. This approach has two significant disadvantages. First, concentration is not uniquely determined by halo mass and redshift---there is significant scatter in concentration at fixed mass and redshift. As such, this scatter will be missing from calculations which depend on halo structure (e.g. galaxy sizes; \citealt{jiang_is_2018}). This scatter cannot be incorporated by simply adding a random perturbation around the median concentration relation as the offset from the median is expected to be correlated across time in any given halo. The second disadvantage of this approach is that it is known that halo structure correlates with the formation history of the halo \citep{navarro_universal_1997,bullock_profiles_2001,ludlow_mass_2013}.

\ 

Based on these correlations, \cite{ludlow_mass-concentration-redshift_2014} (see also \citealt{correa_accretion_2015}) and \cite{ludlow_mass-concentration-redshift_2016} developed models which relate the density profiles of dark matter halos to their assembly histories. The model of \cite{ludlow_mass-concentration-redshift_2016} is based upon the time evolution of the total mass of progenitor halos collapsed (the `collapsed mass history', or CMH), and is able to correctly predict concentrations for both CDM and warm dark matter (WDM) power spectra, while the \cite{ludlow_mass-concentration-redshift_2014} model, which is based on the `mass accretion history' (MAH; the mass of the main progenitor halo) fails to reproduce the concentration mass relation in WDM. As shown by \cite{ludlow_mass-concentration-redshift_2014} and \cite{ludlow_mass-concentration-redshift_2016}, these models work well for halos which are relaxed and in equilibrium, for which concentration is mainly set by formation time and is independent of other factors (such as substructure, recent collapse times, etc.). As such, the models are relevant to a biased set of simulated merger trees, since selecting relaxed halos naturally biases trees to those that form early (on average). While these models were developed and tested primarily on assembly histories extracted from N-body simulations, they were also shown to be applicable to simple, spherical collapse models. As such, they can be similarly applied to assembly histories derived from excursion set-based merger trees. 

It has also been shown \citep[e.g.][]{zehavi_impact_2018} that the formation times of halos correlate with environment. Any model which relates concentration to halo assembly history (and, therefore, formation time) should also consider the relation with environment. In excursion set theory, the large scale environment of a halo has no effect on its assembly history \citep{bond_excursion_1991}, at least in the standard case where the power spectrum is filtered using a window function that is sharp in $k$-space.

In this work we examine the outcome of applying the \cite{ludlow_mass-concentration-redshift_2016} model for halo concentration to merger trees built using the \cite{parkinson_generating_2008} algorithm. We further explore how a simple model for the effects of environment on halo assembly can be introduced into this algorithm, and show that this improves agreement with the distribution of concentrations derived from N-body simulations.

The remainder of this paper is organized as follows. In \S\ref{sec:methods} we describe how we measure the concentrations of N-body halos, and estimate the uncertainties on these measurements. We then describe our algorithm for introducing an environmental dependence into the \cite{parkinson_generating_2008} tree-building algorithm, and how the parameters of this model are constrained. In \S\ref{sec:results} we show the results of this model and compare it to those from N-body simulations. Finally, in \S\ref{sec:discussion} we discuss the implications of our results.

\section{Methods}\label{sec:methods}

Our goal in this work is two-fold: to introduce a model for halo concentrations into the \cite{parkinson_generating_2008} tree-building algorithm which is dependent upon the formation history of each halo, and to compare the results of that model to those from cosmological N-body simulations. This will require determining the full distribution of concentration parameter as a function of halo mass. To that end, we first describe how we measure concentrations of a sample of N-body halos and, importantly, how we determine the uncertainties in these measurements arising from the finite number of particles with which each halo is represented \citep{benson_constraining_2017,benson_mass_2017}.

\subsection{Concentrations of N-body halos}

\subsubsection{Fitting procedure}\label{sec:fitting}

We make use of the \emph{Copernicus Complexio} (``COCO'') simulations of \cite{hellwing_copernicus_2016} to determine the distribution of N-body halo concentrations. As described by \cite{hellwing_copernicus_2016}, halos were identified in the COCO simulations using the friends-of-friends algorithm \citep{davis_evolution_1985}, with a linking length parameter of $b=0.2$. For each halo a mass $M_\mathrm{200c}$ is determined as the mass within a sphere centred on the halo particle with the minimum gravitational potential, and enclosing a mean density equal to 200 times the critical density. As shown by \cite{ludlow_dynamical_2012}, halos of a given mass that collapse very recently (i.e. double their mass in less than a crossing time) have concentrations that are biased relative to the normal concentration-formation time relation. As such, we follow \cite{ludlow_mass-concentration-redshift_2016} and exclude any halos which more than doubled their mass in the last 1.25 crossing times, or approximately the last 3.7~Gyr.

To measure concentrations of halos via their N-body representations we construct histograms of the number of particles in the halo in 31 spherical shells with logarithmically-spaced radii between $3.46\times10^{-2} r_\mathrm{200c}$ and $0.764 r_\mathrm{200c}$, where $r_\mathrm{200c}$ is the radius enclosing a mean density equal to 200 times the critical density, and then fit these using an Einasto profile \citep{einasto_construction_1965}, which has been found to be a good fit to the density profile of cosmological N-body halos \citep{gao_redshift_2008}. In performing the fit we include only those bins which satisfy the inequalities:
\begin{eqnarray}
r &\ge& 2 \epsilon, \\
\kappa(r) &\ge& \kappa_\mathrm{conv},
\end{eqnarray}
where $\epsilon$ is the softening length (230pc for the COCO simulations), and $\kappa(r) = t_\mathrm{relax}(r)/t_\mathrm{circ}(r_\mathrm{200c})$ is a convergence criterion defined by \citeauthor{power_inner_2003}~(\citeyear{power_inner_2003}; their eqn. 20). We choose a value of $\kappa_\mathrm{conv}=1$ to ensure the profile is minimally affected by numerical relaxation. (\citealt{power_inner_2003} advocate $\kappa=0.6$, so we are more conservative in defining convergence as larger $\kappa$ implies better convergence, \citealt{navarro_diversity_2010}). We find that this relaxation criterion exlcudes only 1.2\% of halos. We furthermore retain only those halos for which at least 16 bins satisfy the above inequalities to ensure that the fit is well constrained.

To find the best fit profile for each halo we minimize a goodness-of-fit measure:
\begin{equation}
\phi^2 = \sum_{i=1} \log^2\left({N^\mathrm{(n)}_i\over N^\mathrm{(e)}_i}\right),
\end{equation}
where $N_i^\mathrm{(n)}$ is the number of particles in the $i^\mathrm{th}$ bin of the N-body profile, and $N_i^\mathrm{(e)}$ if the mean number of particles expected in that bin assuming an Einasto profile\footnote{We also considered an alternative approach in which we maximized the Poisson likelihood $\log \mathcal{L} = \sum_i - N^\mathrm{(e)}_i + N^\mathrm{(n)}_i \log N^\mathrm{(e)}_i - \log \Gamma(N^\mathrm{(n)}_i+1)$, which may be expected to be valid if the number of particles in each bin obeys Poisson statistics \protect\citep{benson_constraining_2017}. We found that this leads to small but significant differences in the resulting distribution of concentration parameters. Specifically, the scatter in $\log_{10}c_\mathrm{200c}$ at fixed mass is decreased by around 0.02~dex. However, we find that the Poisson model is not a good description of the distribution of particle number in each bin, presumably because of fluctuations introduced by subhalos and other internal structure. This does, however, point to the need for a more careful understanding of this issue for precision measurements of halo concentrations.}.

Our Einasto profiles are described by three parameters: $m_0$ (the total mass of the halo in units of the nominal $M_\mathrm{200c}$ mass reported by \textsc{subfind}), $r_{-2}$ (the radius at which the logarithmic slope of the density profile equals $-2$, measured in units of the nominal $r_\mathrm{200c}$ radius reported by \textsc{subfind}), and $\alpha$ (the shape parameter of the Einasto profile). We choose to fix $\alpha=0.18$ (consistent with the typical shape of halos reported by \citealt{gao_redshift_2008}) as it is generally not well constrained by the data, and explore broad ranges of the remaining two parameters: $m_0 = (0.5,1.5)$, $r_\mathrm{s}=(0.01,0.50)$. The range for $m_0$ is chosen to be sufficiently broad to encapsulate all plausible profiles, and that on $r_\mathrm{s}$ is chosen to encompass the plausible range of concentrations \citep{gao_redshift_2008}. Once the best-fit parameters for the Einasto profile, $(\hat{m}_0,\hat{r}_{-2})$, have been determined we use them to compute the value of $\hat{r}_\mathrm{200c}$ (i.e. the radius enclosing a mean density equal to 200 times the critical density in the best-fit profile in units of the nominal $r_\mathrm{200c}$ radius reported by \textsc{subfind}, and which may differ slightly from unity) for this profile, and then compute a concentration $c_\mathrm{200c}=\hat{r}_\mathrm{200c}/\hat{r}_{-2}$.

\subsubsection{Uncertainties in N-body concentration estimates}

As has been explored by \cite{trenti_how_2010} and \cite{benson_constraining_2017,benson_mass_2017}, measurements of halo properties from N-body simulations are subject to noise arising from the discrete sampling of the underlying distribution function by a finite number of particles. To explore the effects of the discreteness noise on our concentration measures we perform two Monte Carlo experiments.

In the first, we generate spherically symmetric Einasto density profiles with a range of different concentrations, and sample from each profile using different numbers of particles via a Poisson process \citep{benson_constraining_2017}. These N-body representations are then fit using the procedure described above (without any application of the \cite{power_inner_2003} convergence criterion, which does not apply in this simple experiment). This is repeated for many different random realizations of the particle distribution, and for many different total numbers of particle in the halo to quantify the uncertainty and bias in the recovered concentration.

In our second experiment, we utilize N-body halos from the Millennium Simulation \citep{springel_simulations_2005}, fitting them using the same method as described above. However, we sample particles uniformly at random from each halo (with replacement) to generate a large number of random realizations of the halo, and repeat the fitting procedure on each realization. This sampling is initially done at a rate of $\mu=1$ (where we define $\mu$ as the mean number of times that any given particle will appear in a random realization), and then repeated with lower sampling rates to generate realizations of the halos with fewer particles. This allows us to assess the uncertainty and bias in the measured concentrations for realistic, cosmological halos.

\begin{figure}
\includegraphics[trim={0 0 0 0},clip,width=85mm]{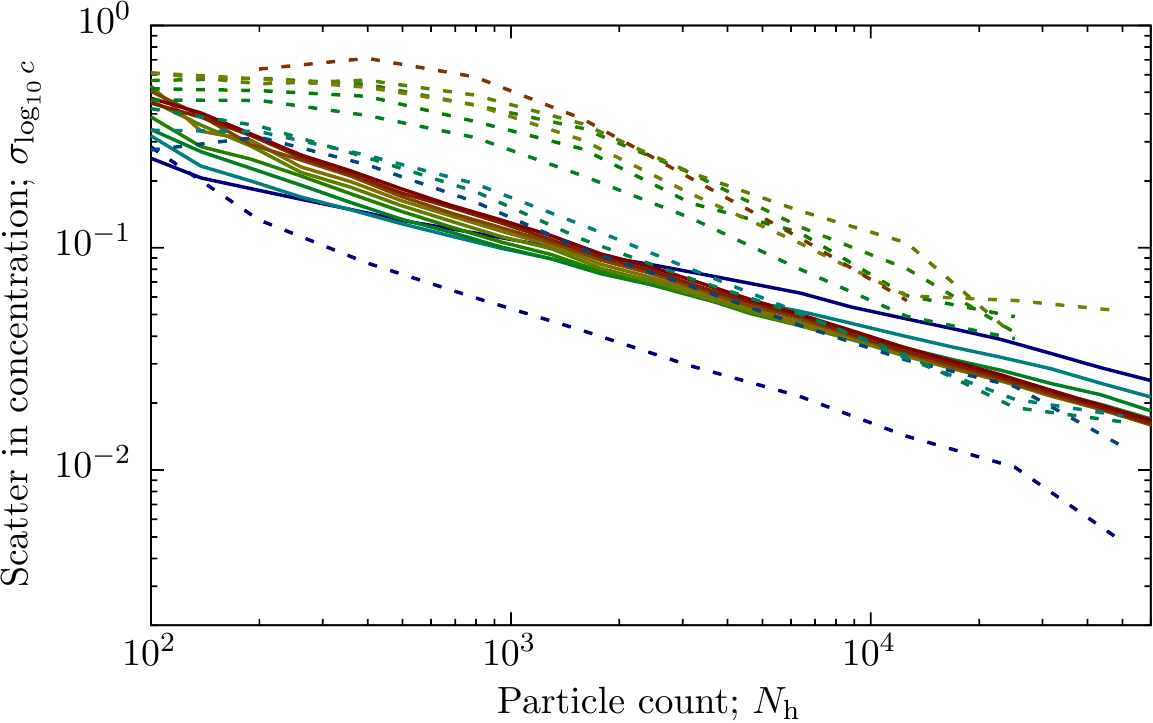}
\caption{Scatter in the concentration parameter, $\log_{10}c_\mathrm{200c}$, derived from Monte Carlo experiments. The scatter, $\sigma_{\log_{10}c_\mathrm{200c}}$, is shown as a function of the number of particles in a halo, $N_\mathrm{h}$, on the $x$-axis, and as a function of concentration (shown by colour---with $c_\mathrm{200c}=4$ shown in blue, increasing to $c_\mathrm{200c}=20$ shown in red). Solid lines show the results of fitting Monte Carlo realizations of idealized, spherical, Einasto profiles. Dashed lines indicate results from fitting cosmological N-body halos, the particles of which were resampled (with replacement) at different rates and fit to estimate concentration. Each dashed line corresponds to the mean scatter measured over all cosmological halos of that concentration. Cosmological halos which deviate significantly from the expected scaling are excluded.}
\label{fig:concentrationErrors}
\end{figure}

The results of these experiments are shown in Figure~\ref{fig:concentrationErrors}. The scatter, $\sigma_{\log_{10}c_\mathrm{200c}}$, is shown as a function of the number of particles in a halo, $N_\mathrm{h}$, on the $x$-axis, and as a function of concentration (shown by colour---with $c_\mathrm{200c}=4$ shown in blue, increasing to $c_\mathrm{200c}=20$ shown in red). Solid lines show the results of fitting Monte Carlo realizations of idealized, spherical, Einasto profiles. Dashed lines indicate results from fitting cosmological N-body halos, with each dashed line corresponding to the mean scatter measured over all cosmological halos of that concentration. Cosmological halos which deviate significantly from the expected scaling are excluded---in these cases the halos are not well-described by Einasto profiles and so we do not expect our fits to give meaningful results.

We fit the dependence of $\sigma_{\log_{10}c_\mathrm{200c}}$ on particle number and concentration using simple polynomial fits. For idealized halos we find that the scatter in concentrations can be described by the model
\idealizedHaloScatterFit
while for cosmological halos the scatter is described by
\cosmologicalHaloScatterFit 
In both cases the dependence of the relation on particle number is consistent with the expected $N_\mathrm{h}^{-1/2}$ scaling due to Poisson sampling. The normalization is significantly higher for the cosmological halos than for the idealized halos.

Achieving an uncertainty in concentration parameter less than 0.1~dex requires $N_\mathrm{h} \gtrsim 2\times10^3$ for idealized halos, and $N_\mathrm{h} \gtrsim 6\times10^3$ for cosmological halos. When fitting our model to match N-body data, we will use the model of equation~(\ref{eq:cosmologicalHaloScatterFit}) to forward model the scatter in concentration parameter.

We also considered the bias in concentration---defined as the difference between the measured and true concentrations. (In the case of cosmological halos, the mean concentration found when fitting to the halo particles sampled at a rate of $\mu=1$ is taken to be our estimate of the true concentration.) Idealized halos show a trend of positive bias in fits, while cosmological halos show no significant trend in bias---the bias estimates for individual halos scatter around zero. As such, we assume no bias when modeling concentrations.

\subsubsection{Concentrations of COCO halos}\label{sec:concentrationFitCOCO}

\begin{figure}
\includegraphics[trim={1mm 0 0 0},clip,width=85mm]{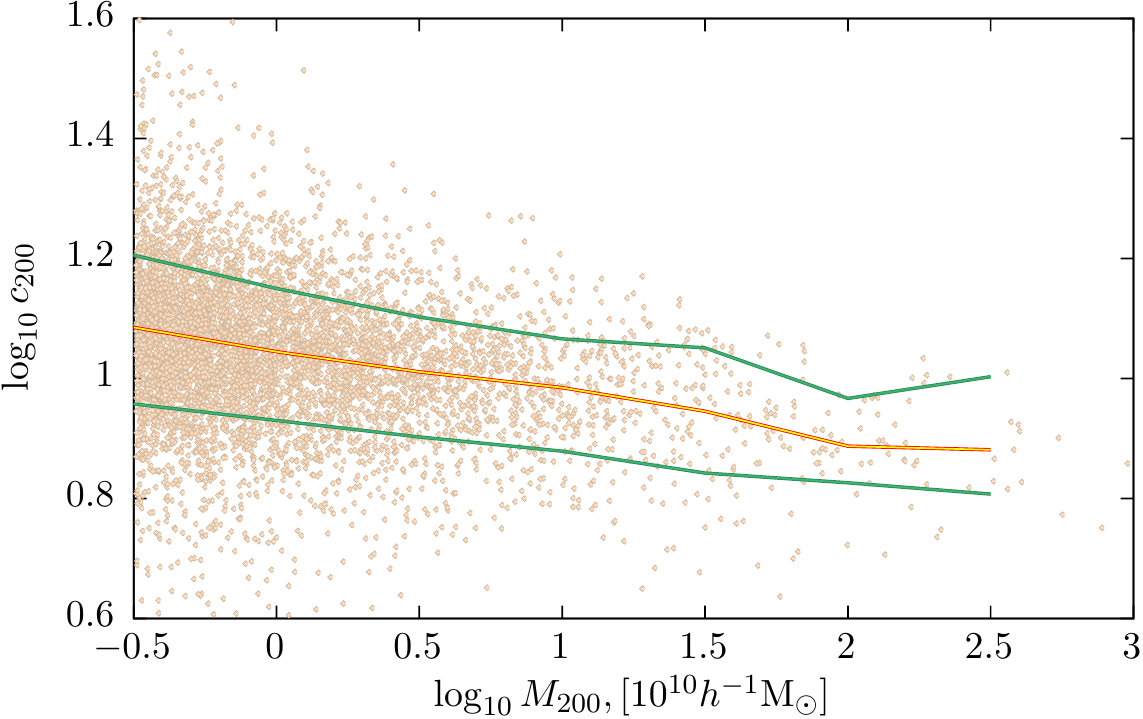}
\caption{Concentrations of halos in the COCO simulation as a function of halo mass. Points show a random subsample of all halos fitted. The solid yellow line shows the median concentration as a function of mass, while the solid green lines show the $16^\mathrm{th}$ and $84^\mathrm{th}$ percentiles of the distribution.}
\label{fig:cocoConcentrations}
\end{figure}

We fit profiles from the COCO simulations \citep{hellwing_copernicus_2016} using the procedure described in \S\ref{sec:fitting}. Results are shown in Figure~\ref{fig:cocoConcentrations}. Points show a random subsample of all halos fitted. The solid yellow line shows the median concentration as a function of mass, while the solid green lines show the $16^\mathrm{th}$ and $84^\mathrm{th}$ percentiles of the distribution. The expected scatter in concentration arising from finite particle number effects (as computed using the model of equation~\ref{eq:cosmologicalHaloScatterFit}) is negligibly small, less than $0.035$~dex, in this figure as all halos shown have $N_\mathrm{p}>25,000$ particles.

\subsection{Merger trees}\label{sec:trees}

Our goal is to apply the concentration model of \cite{ludlow_mass-concentration-redshift_2016} to merger trees generated using an excursion set-based approach. Specifically, we will use the algorithm of \cite{parkinson_generating_2008}---which is itself a modification of the algorithm of \cite{cole_hierarchical_2000}---to construct merger trees. In Appendix~\ref{sec:cole2000Optimize} we describe a small optimization to the tree building algorithm of \cite{cole_hierarchical_2000}.

\subsubsection{Concentrations}

To compute concentrations of our merger trees we apply either the \cite{ludlow_mass-concentration-redshift_2014} or \cite{ludlow_mass-concentration-redshift_2016} algorithms. In the \cite{ludlow_mass-concentration-redshift_2016} algorithm the mean density within the scale radius, $\langle \rho_{-2} \rangle$, for a halo is given by
\begin{equation}
  \frac{\langle \rho_{-2} \rangle}{\rho_0} = C \frac{\rho_\mathrm{crit}(t_\mathrm{c})}{ \rho_0},
  \label{eq:ludlow16}
\end{equation}
where $t_\mathrm{c}$ is the time at which the halo had first assembled a mass $M_{-2}=4\pi\langle \rho_{-2} \rangle r_{-2}^3/3$ into progenitors of mass greater than $f M$, where $M$ is the mass of the halo in question. To find $r_{-2}$ for any given halo in a merger tree we use an iterative approach. We make an initial guess for $r_{-2}$ (typically based on the median concentration-mass-redshift relation) and use this, together with the known mass and virial radius of the halo, to determine $M_{-2}$. We then search backward in time through the progenitors of the halo until we find the time, $t_\mathrm{c}$, at which the halo first had a total mass equal to $M_{-2}$ in progenitor halos of mass $f M$ or greater. An estimate of $\langle \rho_{-2} \rangle$ can then be made from eqn.~(\ref{eq:ludlow16}), which in turn can be used (along with the mass and virial radius of the halo) to compute an updated estimate of $r_{-2}$. This new value of $r_{-2}$ is used in the next iteration of this procedure, which is continued until the value of $r_{-2}$ has converged to a desired level. In this way, scale radii, $r_{-2}$, are found for all halos in the merger tree\footnote{For halos close to the resolution limit of our merger trees progenitors of mass $f M$ may not be resolved. In such cases we can simply assign a concentration from a measured concentration-mass-redshift relation. In this work we do not care about such halos as we ensure that halos for which we wish to compute the concentration are sufficiently well resolved that this problem does not occur.}.

For the algorithm of \cite{ludlow_mass-concentration-redshift_2014} we use a similar approach, except that the condition that the mass in all progenitors above mass $f M$ must equal $M_{-2}$ at $t_\mathrm{c}$, we instead require that the mass in the primary (most massive) progenitor equals $M_{-2}$.

\subsubsection{Environmental dependence}

Since we wish to examine the effects of environment on our merger trees, we select for each tree an environmental overdensity (defined within some spherical region of radius $R_\mathrm{e}$) in the linear regime extrapolated to the present epoch, $\delta_\mathrm{e}$. This is drawn from the distribution of \citeauthor{mo_analytic_1996}~(\citeyear{mo_analytic_1996}; their eqn. 9), that is from a normal distribution conditioned upon the fact that the region has not exceeded the threshold for collapse on any larger scale. This environmental overdensity is assumed to be the same for all progenitor halos in the given merger tree.

Each merger tree, with final halo of mass $M_\mathrm{i}$ (defined such that $M_\mathrm{i} > M_\mathrm{i-1}$), is considered to have an abundance
\begin{equation}
n_\mathrm{i} = \frac{1}{M_\mathrm{i}} \int_{\sqrt{M_\mathrm{i-1}M_\mathrm{i}}}^{\sqrt{M_\mathrm{i}M_\mathrm{i+1}}} M n(M,\delta_\mathrm{e}) \mathrm{d}M.
\end{equation}
That is, the abundance is chosen such that the number of such trees per unit volume contains the total mass expected from the halo mass function $n(M,\delta_\mathrm{e})$ within the mass interval associated with the tree.

To determine the environment-dependent halo mass function, $n(M,\delta_\mathrm{e})$, we use the peak-background split approach \citep{bond_excursion_1991,bond_peak-patch_1996}. This is formally derived within the excursion set theory, and predicts a mass function of the \cite{press_formation_1974} form but with the mappings $\delta_\mathrm{c} \rightarrow \delta_\mathrm{c} - \delta_\mathrm{e}$ and $\sigma^2(M) \rightarrow \sigma^2(M) - \sigma^2(M_\mathrm{e})$, where $M_\mathrm{e}$ is the mass corresponding to the spherical region used to define the environment, and $\sigma^2(M)$ is the fractional variance in the mass in spheres containing mean mass $M$ in the linear theory density field extrapolated to the present day. In this work, we use the mass function of \cite{sheth_excursion_2002} with parameter values tuned as described in \S\ref{sec:calibration} rather than the Press-Schechter mass function. We retain this same mapping of $\delta_\mathrm{c}$ and $\sigma^2(M)$ suggested by the peak-background split.

In the case of the Press-Schechter mass function, when the peak-background split mass function is integrated over all environments (weighted by the appropriate distribution function---see \citealt{mo_analytic_1996}) the result is the original Press-Schechter mass function (i.e. that unconditioned on environment). For halo masses which exceed the mass of the background, the unconditioned mass function applies. Therefore, in the case of the Press-Schechter mass function, the environmentally-averaged mass function is identical to the unconditioned mass function.

The same is not true for the Sheth-Tormen mass function. Instead, we find that averaging this mass function over environment leads to a result slightly larger than the corresponding unconditioned mass function. This overshoot peaks at around 5\% close to the background mass, and approaches zero at low masses. This overshoot occurs because the distribution function we use for the background overdensity is derived for the constant barrier corresponding to the Press-Schechter mass function. In principle the correct distribution function could be found for the curved barrier corresponding to the Sheth-Tormen mass function, by numerical solution of the barrier crossing problem. Given the small magnitude of the overshoot we do not seek to find the correct background distribution here, but instead simply recalibrate the parameters of the Sheth-Tormen mass function such that when environmentally-averaged it agrees well with N-body measures of the mass function. Because of the overshoot, there is a discontinuity in the mass function at $M_\mathrm{e}$. To correct for this we multiply the mass function for $M>M_\mathrm{e}$ by a fixed factor to remove the discontinuity.

While the peak-background split can be used to accurately predict the environmental dependence of the halo mass function, it predicts no environmental dependence in halo formation histories. This is easy to see in excursion set theory, as the behaviour of the trajectory on scales larger than the collapse scale (at variances smaller than the collapse variance) is uncorrelated with the behaviour of the trajectory on smaller scales (larger variances)\footnote{This is true under the usual assumption of a sharp-$k$ filter---for other filters the trajectory is non-Markovian \protect\citep{maggiore_halo_2010}.}.

In the merger tree algorithm of \cite{parkinson_generating_2008} the quantity $w(t)=\delta_\mathrm{c}(t)/D(t)$, where $\delta_\mathrm{c}$ is the critical linear theory overdensity for collapse at time $t$ and $D(t)$ is the linear theory growth factor, plays the role of a ``time'' variable, and sets the threshold which perturbations must reach to collapse and form halos. Therefore, to introduce an environmental dependence into this algorithm, we make an empirical modification such that
\begin{equation}
w(t;\delta_\mathrm{e}) = \delta_\mathrm{c}(t) D^{\alpha\delta_\mathrm{e}-1}(t),
\end{equation}
where $\alpha$ is a parameter to be determined. This introduces a dependence on the mapping between merger tree ``time'', $w(t)$, and true cosmic time, $t$, which depends on the environmental overdensity.

The choice of the radius used to define environment, $R_\mathrm{e}$, is somewhat arbitrary, but may influence the value of $\alpha$ in our model. We choose to use $R_\mathrm{e}=5 h^{-1}$~Mpc (where $h=H_0/100\hbox{km~s}^{-1}\hbox{Mpc}^{-1}$) as this is larger than the collapse scale for the vast majority of halos, while still being a relatively ``local'' measure of environment. This point is discussed further in \S\ref{sec:formation}.

\subsubsection{Calibration}\label{sec:calibration}

Several predictions of our model---specifically the environment-averaged halo mass function, halo progenitor mass functions, and the distribution of halo concentrations as a function of halo mass---will be altered as a consequence of our introduction of environmental dependences into both the halo mass function, $n(M;\delta_\mathrm{e})$, by adopting the peak-background split model, and the halo collapse threshold, $w(t;\delta_\mathrm{e})$. Therefore, we recalibrate the parameters of these models to match measurements from N-body simulations. This calibration is carried out by running a \MCMC\ simulation. Our approach follows that of \cite{benson_mass_2017} in detail, including utilizing the same \MCMC\ algorithm and convergence criteria. \cite{benson_mass_2017} constrained the parameters of the \cite{sheth_excursion_2002} mass function, and the \cite{parkinson_generating_2008} merger tree algorithm to match the mass function and progenitor mass functions measured from the \MDPL\ simulation \citep{klypin_multidark_2016}, after having removed splashback halos to avoid double-counting. We constrain these parameters in the same way here, except that we additionally average halo and progenitor mass functions over environment, and include the parameter $\alpha$ as an additional parameter in our \MCMC\ simulation.  Furthermore, when constructing progenitor mass functions we adopt the error distribution of \cite{trenti_how_2010} when convolving the intrinsic halo mass function with the expected error distribution. As the \cite{trenti_how_2010} errors are dominated by the effects of missing structure below the resolution limit, we expect that errors in halo masses will be correlated across time. This correlation should be measured directly from N-body simulations, using a methodology similar to that adopted by \cite{trenti_how_2010}. Lacking such an analysis, we instead adopt a simple model to describe the covariance between the halo masses of parent and progenitor halos. Given a parent halo, ``0'', and progenitor halo, ``1'' we assume that their masses are drawn from a distribution with covariance matrix $\mathbfss{S}$. The diagonal elements of this matrix are set to $\sigma^2_0$ and $\sigma^2_1$, where $\sigma_0$ and $\sigma_1$ are the root-variances in the parent and progenitor halo mass according to \cite{trenti_how_2010}. For the off-diagonal elements we assume that the covariance is $C \sigma_0 \sigma_1$, with
\begin{equation}
 C = C_0 \left({M_1\over M_0}\right)^{C_\mathrm{m}} \left({a_1\over a_0}\right)^{C_\mathrm{a}},
\end{equation}
where $a$ is expansion factor, $M$ is halo mass, and $C_0$, $C_\mathrm{m}$, and $C_\mathrm{a}$ are parameters to be determined. We include these three nuisance parameters in our MCMC simulation, adopting broad, uniform priors in the interval 0--1 for $C_0$ and 0--2 for both $C_\mathrm{m}$, and $C_\mathrm{a}$, and will marginalize over them in our final analysis.

The concentration model of \cite{ludlow_mass-concentration-redshift_2016} has two free parameters, $f$ and $C$. We include the parameters $C$ and $f$ in our \MCMC\ simulation, and constrain the model to match the distribution of concentrations found in the COCO simulations (see \S\ref{sec:fitting}). Specifically, we construct histograms of the concentrations of halos in seven logarithmically-spaced mass bins spanning the range $9.4 < \log_{10}(M/M_\odot) < 12.4$, and adopt a log-likelihood for each mass bin of
\begin{equation}
\log\mathcal{L} = -\frac{1}{2} \Delta \mathbfss{C}^{-1} \Delta^{\rm T},
\end{equation}
where $\Delta$ is a vector of differences between the N-body and model concentration histograms, and $\mathbfss{C}$ is a covariance matrix. The covariance matrix is taken to be the sum of that of the N-body halo histogram, and that of the model halo histogram, both assuming Poisson counting statistics. When constructing the histogram of model halo concentrations we exclude halos which have more than doubled their mass in the last 1.25 crossing times (3.7~Gyr; see \S\ref{sec:fitting}; which we find excludes 2.6\% of halos), and smooth by a Gaussian with width chosen to match the expected uncertainty in N-body halo concentration estimates (see eqn.~\ref{eq:cosmologicalHaloScatterFit})---this smoothing is taken into accounting when computing the covariance matrix.

For the parameters of the \cite{sheth_excursion_2002} halo mass function and \cite{parkinson_generating_2008} merger tree algorithm we adopt the same priors as were used by \cite{benson_mass_2017}. For the parameter $\alpha$ we adopt a uniform prior in the range $0.0$ to $0.4$---we expect this parameter to be positive to induce early formation times in higher density environments (see \S\ref{sec:results}), and we find that values larger than $0.4$ can lead to $w(t;\delta_\mathrm{e})$ becoming an increasing function of $t$, which would result in progenitor halos collapsing \emph{after} their descendants and so is physically impossible. For the parameters $C$ and $f$ we adopt uniform priors of $(100,800)$ and $(0.01,0.10)$ respectively. These priors are broad and include the values found by \cite{ludlow_mass-concentration-redshift_2016} to match results for both cosmological N-body halos, $(C,f)=(400,0.02)$, and simple spherical collapse models, $(C,f)=(650,0.02)$. Furthermore, \cite{ludlow_mass-concentration-redshift_2016} find that $C$ and $f$ correlate with the slope of the $\langle \rho_{-2}\rangle$--$\rho_\mathrm{crit}(z_{-2})$ relation, when $f$ approaches the mass $M_{-2}/M_{200}$ (which is typically $0.1$--$0.2$). This correlation is currently neglected in our modeling, which further motivates our choice to restrict $f$ to values less than $0.1$.
f restricted to small values, e.g. f=(0.01,0.1)

\section{Results}\label{sec:results}

\begin{table}
  \caption{Maximum posterior probability values and the region containing 68\% of the posterior probability for all parameters used in our model (the nuisance parameters $C_0$, $C_\mathrm{m}$ and $C_\mathrm{a}$ are not listed). For each parameter we also list the previously determined value where available.}
  \label{tb:constraints}
  \bgroup
  \def\arraystretch{1.5}
\begin{tabular}{llll}
\hline
           & \textbf{This work}        & \textbf{Previous work}                                                              \\
\hline
$a$        & \parameterBestFita        & $+0.874^{+0.005}_{-0.005}$ & \protect\cite{benson_mass_2017}                        \\
$p$        & \parameterBestFitp        & $-0.031^{+0.005}_{-0.005}$ & \protect\cite{benson_mass_2017}                        \\
$A$        & \parameterBestFitA        & $+0.332^{+0.0002}_{-0.0002}$ & \protect\cite{benson_mass_2017}                        \\
$G_0$      & \parameterBestFitGZero    & $+0.635^{+0.011}_{-0.0002}$ & \protect\cite{benson_mass_2017}                        \\
$\gamma_1$ & \parameterBestFitgammaOne & $+0.176^{+0.002}_{-0.015}$ & \protect\cite{benson_mass_2017}                        \\
$\gamma_2$ & \parameterBestFitgammaTwo & $+0.041^{+0.001}_{-0.009}$ & \protect\cite{benson_mass_2017}                        \\
$\alpha$   & \parameterBestFitalpha    & ---                                                                                 \\
$C$        & \parameterBestFitC        & $+400.0$                  & \protect\cite{ludlow_mass-concentration-redshift_2016} \\
$f$        & \parameterBestFitf        & $+0.020$                  & \protect\cite{ludlow_mass-concentration-redshift_2016} \\
\hline
\end{tabular}
\egroup
\end{table}

The parameter constraints derived from our \MCMC\ simulation are listed in Table~\ref{tb:constraints}, along with previously determined values. The maximum likelihood values of the \cite{sheth_excursion_2002} and \cite{parkinson_generating_2008} model parameters are significantly shifted by the inclusion of environmental dependences, but the resulting halo and progenitor mass functions remain equally good matches to the \MDPL\ N-body measurements. We find that the parameter $\alpha$ is strongly constrained to be non-zero, indicating that the N-body data are better fit by a model with environmental dependence in halo merger rates. For the parameters $(C,f)$ of the \cite{ludlow_mass-concentration-redshift_2016} model we find best-fit values higher than those reported by \cite{ludlow_mass-concentration-redshift_2016}. This may reflect some difference in the structure of our merger trees compared to those extracted from N-body simulations, but we note that the posterior distribution over $(C,f)$ shows that these two parameters are strongly degenerate in the direction which includes the $(C,f)=(400,0.02)$ results favoured by \cite{ludlow_mass-concentration-redshift_2016}.

\subsection{Concentrations}

\begin{figure*}
  \begin{tabular}{cc}
   \includegraphics[trim={0mm 0 0mm 0},clip,width=85mm]{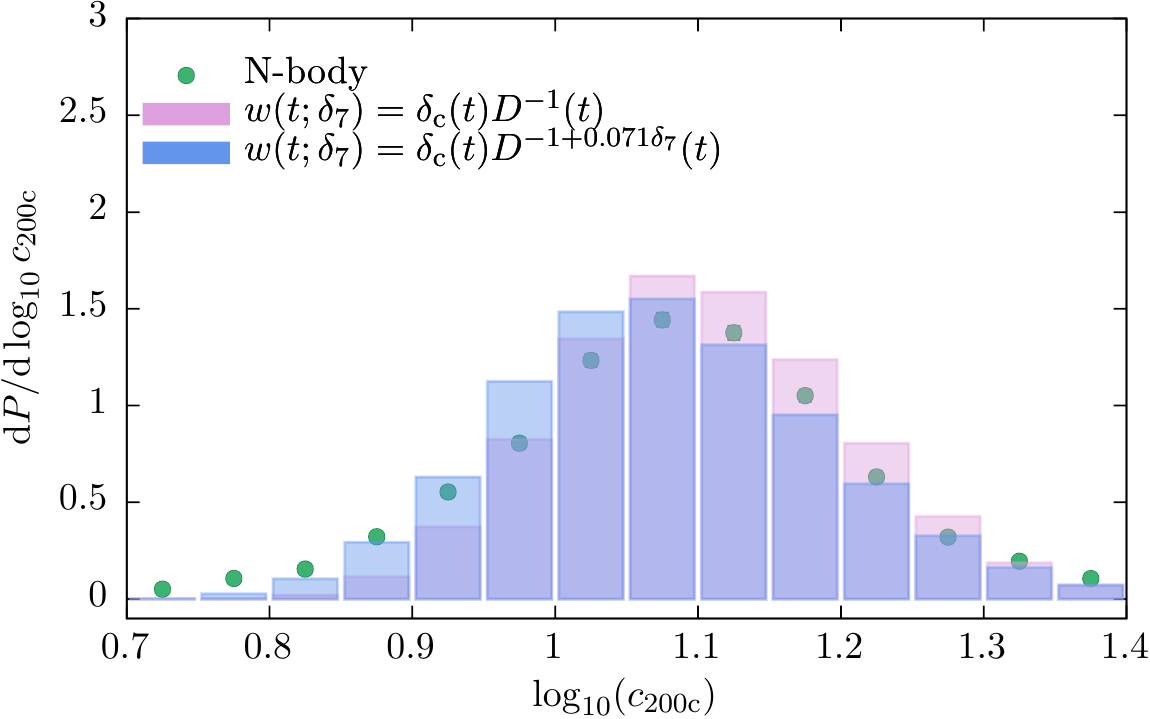} & \includegraphics[trim={0mm 0 0mm 0},clip,width=85mm]{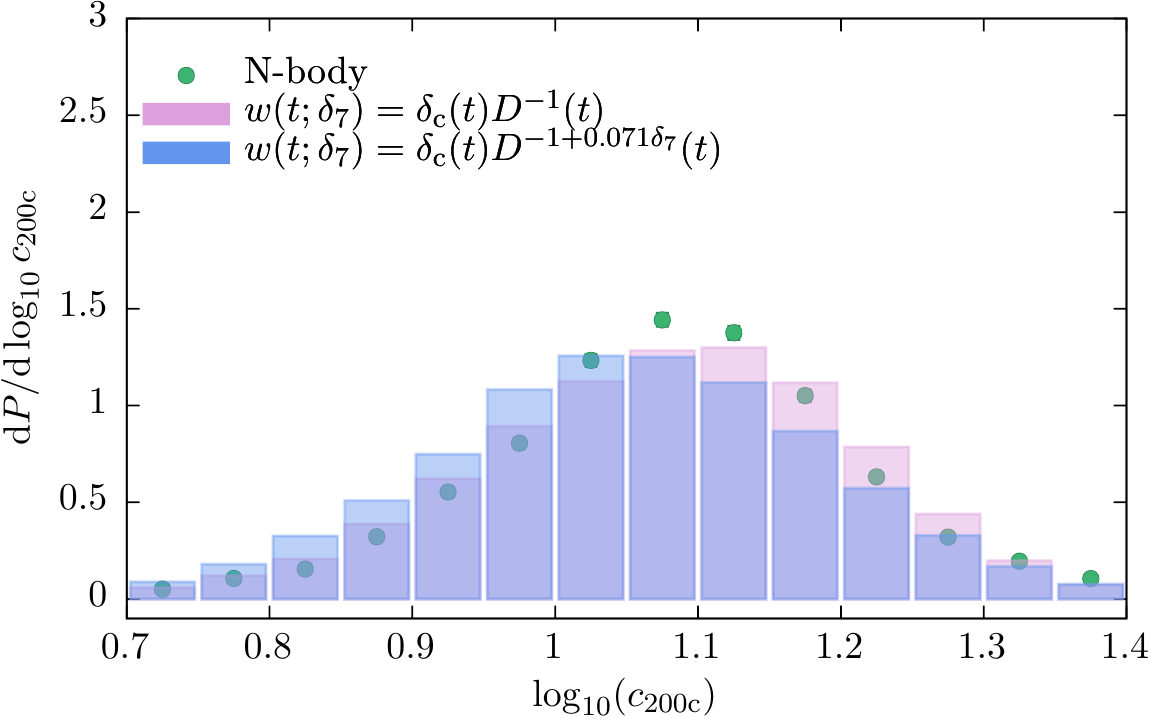}
  \end{tabular}
  \caption{Distribution of concentration, $c_\mathrm{200c}$, for halos in the mass range $\ludlowSixteenNBodyMassMinimum\le \log_{10} (M_\mathrm{200c}/\mathrm{M}_\odot) < \ludlowSixteenNBodyMassMaximum$ at $z=0$. Green points show results for concentration measured directly from N-body halos (see \S\ref{sec:concentrationFitCOCO}). Histograms show results obtained using the models of \protect\cite{ludlow_mass-concentration-redshift_2016} (left panel) and \protect\cite{ludlow_mass-concentration-redshift_2014} (right panel). Pink histograms indicate results when the collapse threshold, $w(t)$, is independent of environment, while blue histograms indicate results when $w(t)$ scales with environmental density as indicated in the panels.}
  \label{fig:concentrationDistribution}
\end{figure*}

\begin{table}
  \caption{Mean and scatter in $\log_{10}c_\mathrm{200c}$ for different concentration models, and as measured from COCO N-body halos, for halos in the mass range $\ludlowSixteenNBodyMassMinimum\le \log_{10} (M_\mathrm{200c}/\mathrm{M}_\odot) < \ludlowSixteenNBodyMassMaximum$ at $z=0$.}
  \label{tbl:concentrationDistribution}
  \begin{center}
    \begin{tabular}{lll}
      \hline
      \textbf{Model} & \boldmath{$\langle \log_{10}c_\mathrm{200c} \rangle$} & \boldmath{$\sigma_{\log_{10}c_\mathrm{200c}}$} \\
      \hline
      COCO (N-body) & \ludlowSixteenNBodyConcentrationMean & \ludlowSixteenNBodyConcentrationScatter \\
      Ludlow et al. (2016), $\alpha=\ludlowSixteenNoEnvironmentExponentCollapseThreshold$ & \ludlowSixteenNoEnvironmentConcentrationMean  & \ludlowSixteenNoEnvironmentConcentrationScatter  \\
      Ludlow et al. (2016), $\alpha=\ludlowSixteenEnvironmentExponentCollapseThreshold$ & \ludlowSixteenEnvironmentConcentrationMean    & \ludlowSixteenEnvironmentConcentrationScatter    \\
      Ludlow et al. (2014), $\alpha=\ludlowFourteenNoEnvironmentExponentCollapseThreshold$ & \ludlowFourteenNoEnvironmentConcentrationMean & \ludlowFourteenNoEnvironmentConcentrationScatter \\
      Ludlow et al. (2014), $\alpha=\ludlowFourteenEnvironmentExponentCollapseThreshold$ & \ludlowFourteenEnvironmentConcentrationMean   & \ludlowFourteenEnvironmentConcentrationScatter   \\
      \hline
    \end{tabular}
  \end{center}
\end{table}

Figure~\ref{fig:concentrationDistribution} shows the distribution of concentrations, $c_\mathrm{200c}$, for halos in the mass range $\ludlowSixteenNBodyMassMinimum\le \log_{10} (M_\mathrm{200c}/\mathrm{M}_\odot) < \ludlowSixteenNBodyMassMaximum$ at $z=0$. The model of \cite{ludlow_mass-concentration-redshift_2014} (right panel) produces a distribution which matches N-body results (shown by green points) quite well when no dependence on environmental density is included in $w(t;\delta_7)$. However, as shown by \cite{ludlow_mass-concentration-redshift_2016}, the \cite{ludlow_mass-concentration-redshift_2014} has significant failings, such as failing to reproduce the concentrations of halos in WDM models. The model of \cite{ludlow_mass-concentration-redshift_2016} (left panel) predicts a distribution which is too narrow when $w(t)$ is independent of environmental density, but matches the N-body results reasonably well when $w(t;\delta_7)=\delta_\mathrm{c}(t)D^{\ludlowSixteenEnvironmentExponentCollapseThreshold \delta_7-1}(t)$. Table~\ref{tbl:concentrationDistribution} summarizes the mean and scatter in concentration in a narrow range of halo mass as measured from the COCO N-body halos, and predicted by the \cite{ludlow_mass-concentration-redshift_2014} and \cite{ludlow_mass-concentration-redshift_2016} models applied to our merger trees. Even with an environmental dependence introduced into $w(t;\delta_7)$ the \cite{ludlow_mass-concentration-redshift_2016} model is unable to produce a scatter quite as large as that measured for N-body halos. This is largely driven by the tail of low-concentration halos seen in the N-body simulation which is not matched by the \cite{ludlow_mass-concentration-redshift_2016} model applied to our excursion set-derived merger trees.

\begin{figure}
  \includegraphics[trim={0 0 0 0},clip,width=85mm]{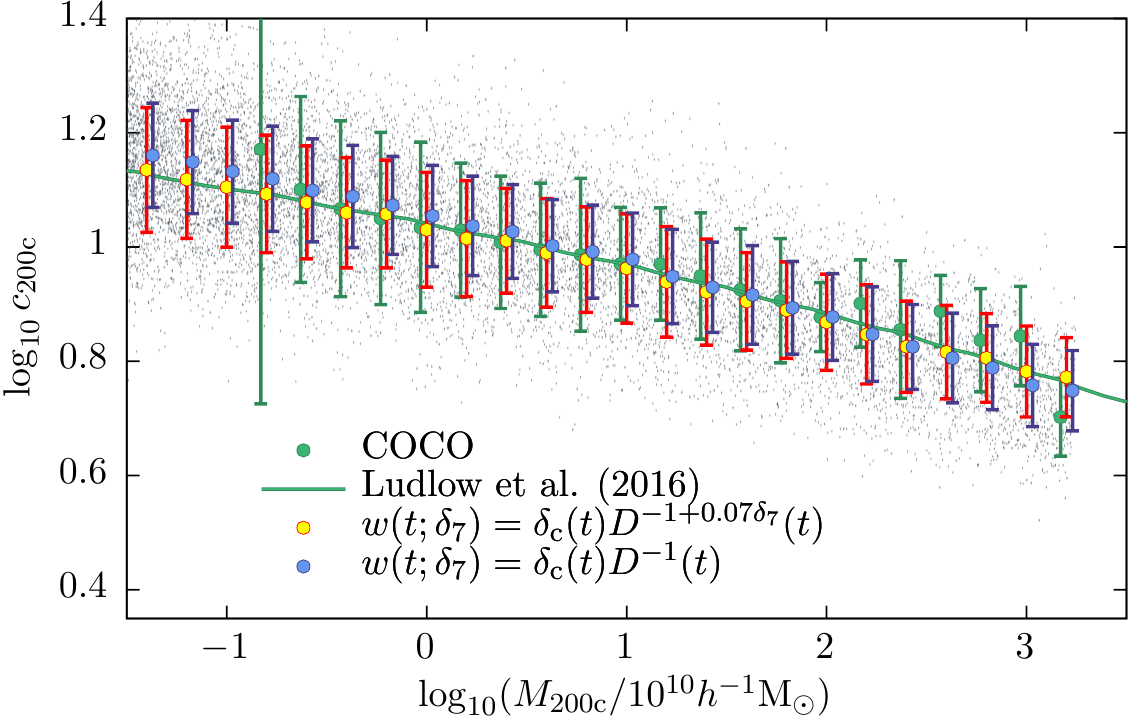}
  \caption{The concentration--mass relation. Points indicate the mean $\log_{10}c_\mathrm{200c}$, while error bars show the root variance in $\log_{10}c_\mathrm{200c}$. Green points are measured from dark matter halo profiles in the COCO-COLD N-body simulation (see \S\ref{sec:concentrationFitCOCO}), while yellow and blue points are from this work using the algorithm of \protect\cite{ludlow_mass-concentration-redshift_2016}, with and without an environmental dependence in $w(t)$ (as shown in the figure). Small grey points show individual halos from this work with $w(t;\delta_7)=\delta_\mathrm{c}(t)D^{\ludlowSixteenEnvironmentExponentCollapseThreshold \delta_7-1}(t)$. The green line indicates the concentration--mass relation computed by \protect\cite{ludlow_mass-concentration-redshift_2016}.}
  \label{fig:concentrationMassRelation}
\end{figure}

Figure~\ref{fig:concentrationMassRelation} shows the concentration--mass relation as measured from N-body halo profiles, and as predicted by the model of \cite{ludlow_mass-concentration-redshift_2016}.

\subsection{Formation epochs}\label{sec:formation}

\begin{figure}
  \includegraphics[trim={0 0 0 0},clip,width=85mm]{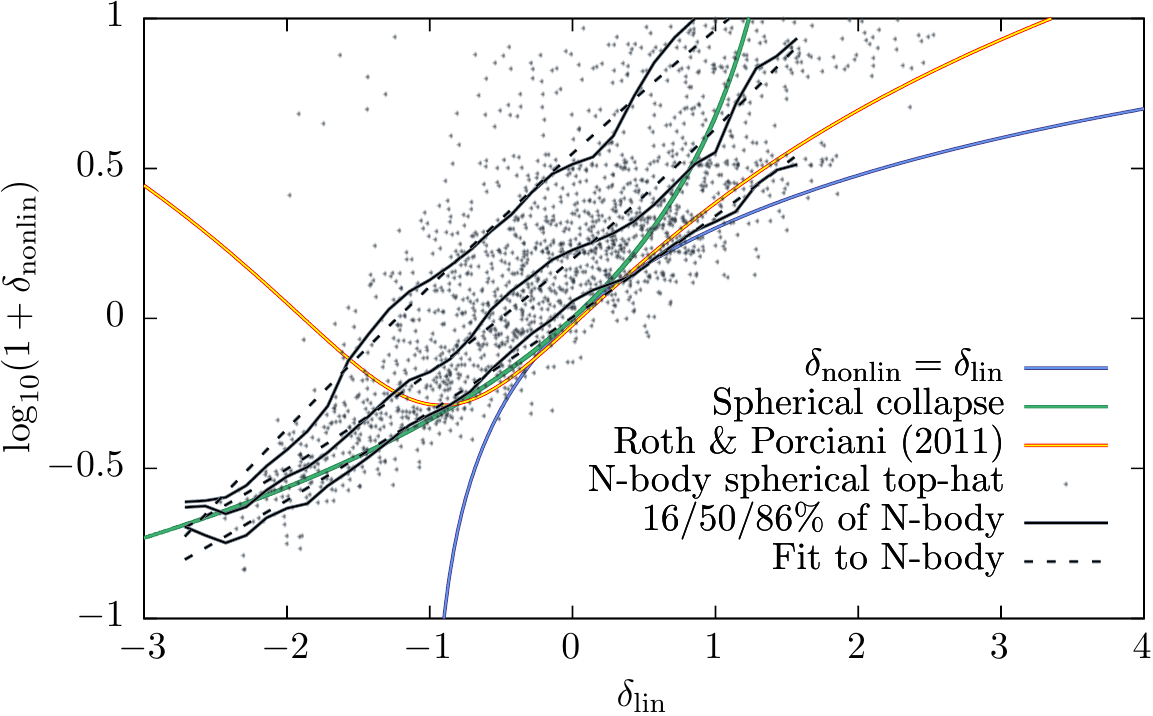}
  \caption{Mapping from Lagrangian, linear overdensity to non-linear Eulerian overdensity. The blue line shows the one-to-one relation, the green curve shows the expectation from spherical collapse, while the yellow line shows the fitting function of \protect\cite{roth_testing_2011}. Grey points show measurements of overdensity around individual particles in an N-body simulation. Solid black lines show the 16/50/84 percentiles of the distribution of these points, while the dashed black lines show polynomial fits to these percentiles.}
  \label{fig:environmentLinearNonLinearRelation}
\end{figure}

N-body halos show a correlation between environmental overdensity and formation time (defined as the time at which a given halo had first assembled 50\% of its final mass into a single progenitor halo) as shown, for example, by \cite{zehavi_impact_2018}. While our model of environmental dependence was constructed to explain the scatter in halo concentrations at fixed mass, we can also ask whether it produces the correct correlation between environment and formation time. 

To assess this correlation, and to compare to the measurements of \cite{zehavi_impact_2018}, we must account for the fact that our environmental overdensity, $\delta_\mathrm{e}$, is the linear theory overdensity in the Lagrangian volume surrounding each halo, while that measured by \cite{zehavi_impact_2018} is the non-linear overdensity in an Eulerian volume surrounding each halo. While a fitting function relating these quantities has been proposed by \cite{roth_testing_2011} we find that it does not well describe the relation here as we are interested in the relation for special points in the density field, namely the locations of halos. We therefore measure the linear theory Lagrangian, and non-linear Eulerian overdensities around particles in the Millennium N-body simulation (as was used by \citealt{zehavi_impact_2018}).

\cite{zehavi_impact_2018} define environment based on a $5 h^{-1}$~Mpc Gaussian smoothing of the N-body particle field (computed in $2 h^{-1}$~Mpc cells), which they label $\delta_5$. Based on their Figure~2, the median value of $1+\delta_5$ is around 2 for most halo masses. As such, these regions have collapsed in radius by a factor of approximately $2^{1/3}$ relative to their initial, Lagrangian volume. We therefore choose to use a radius $R_\mathrm{e}=2^{1/3} 5h^{-1}\,\hbox{Mpc}=6.3h^{-1}\,\hbox{Mpc}$ to define environment in our model\footnote{We did not alter the value of $\alpha$ used in constructing these merger trees, even though it was constrained using $R_\mathrm{e}=5h^{-1}$~Mpc rather than $R_\mathrm{e}=6.3h^{-1}$~Mpc. Since the dispersion in $\delta_\mathrm{e}$ will depend on $R_\mathrm{e}$ it is possible that $\alpha$ should be recalibrated for each $R_\mathrm{e}$. Alternatively, it is possible that one specific value of $R_\mathrm{e}$ is the optimal choice (given some suitable metric to judge optimality) for our model and should always be used to define environment (in which case comparisons to measures using a different definition of environment become more difficult). We leave a full investigation of these issues to a future work.}.

Using the particle distribution in the Millennium simulation we measure the Lagrangian overdensity in spheres of radius $R_\mathrm{e}$ in the initial conditions around each halo's most-bound particle, and extrapolate this to $z=0$ assuming linear perturbation theory. We then determine $\delta_5$ in the $z=0$ density field for the same particles. The results are shown in Fig.~\ref{fig:environmentLinearNonLinearRelation}. We model the distribution of non-linear, Eulerian overdensity at each linear, Lagrangian overdensity as a log-normal distribution, and determine the mean and dispersion of this distribution from the measured points. Then, for each halo in our model, we take its assigned linear, Lagrangian overdensity, and draw a non-linear, Eulerian overdensity from the appropriate log-normal distribution.

\begin{figure}
  \includegraphics[trim={1mm 0 0mm 0},clip,width=85mm]{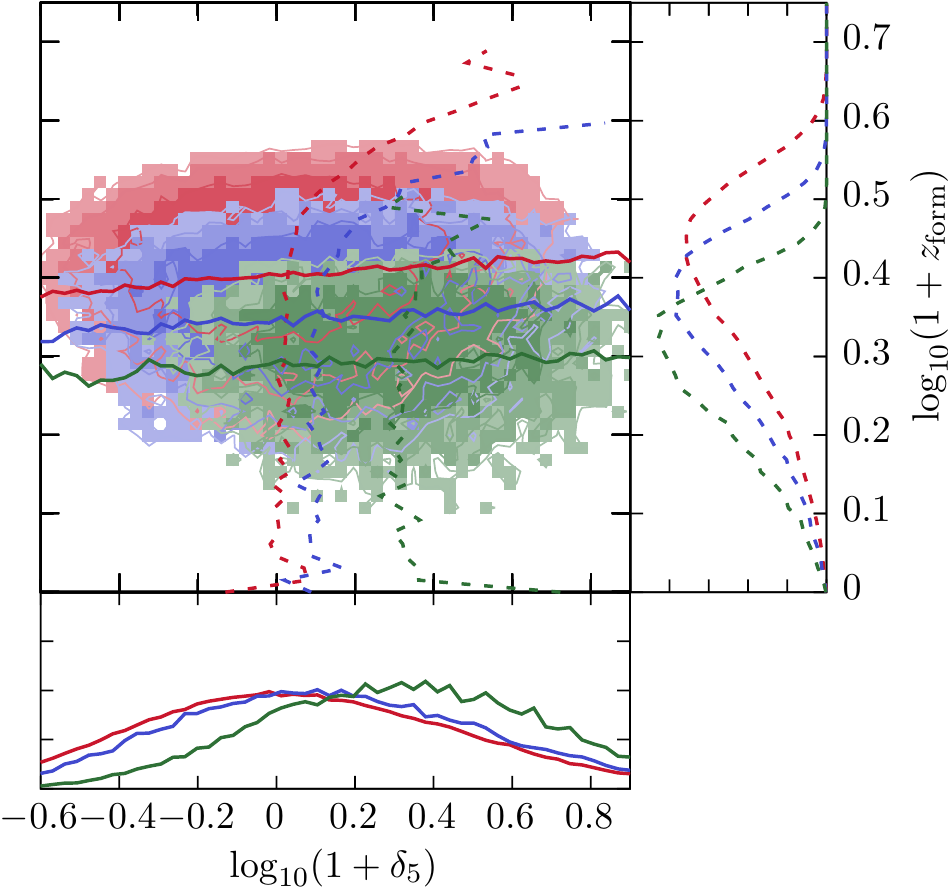}
  \caption{Joint and marginalized distributions of non-linear, Eulerian density contrast and formation epoch for halos in three narrow ranges of $\log_{10}(M_\mathrm{FoF}/h^{-1}\mathrm{M}_\odot)$: 11.0--11.2 (red), 12.0--12.2 (blue), and 13.0--13.2 (green). Formation epoch is defined as the epoch at which the primary progenitor of the $z=0$ halo first reaches half of the final mass. Overlaid on the joint distribution are medians of formation epoch (solid lines) and of density contrast (dashed lines). This figure is intended to be compared directly to that of \protect\cite{zehavi_impact_2018}.}
  \label{fig:environmentFormation}
\end{figure}

Figure~\ref{fig:environmentFormation} shows joint and marginalized distributions of density contrast and formation epoch in three different halo mass ranges, and can be compared directly to Fig.~2 of \cite{zehavi_impact_2018} in which these same distributions were measured directly from the Millennium Simulation \citep{springel_simulations_2005}. To construct these distributions from our model we generated large samples of merger trees as described in \S\ref{sec:trees}, matching the cosmological parameters and power spectrum of the Millennium Simulation. The resulting distributions of overdensity and formation epoch are in qualitative agreement with those found by \cite{zehavi_impact_2018}. In particular, the marginalized distributions of overdensity agree well with \cite{zehavi_impact_2018}, and are in good quantitative agreement for the position of the mode of the distribution, and the width. Marginalized distributions for formation epoch also agree well. Finally, the joint distribution shows a correlation between formation epoch and environmental overdensity which agrees moderately well with that found by \cite{zehavi_impact_2018}---without our overdensity-dependent modification of the collapse threshold no correlation would exist. A more quantitative comparison of the effects of environment on assembly history in excursion set and N-body merger trees is currently limited by the different definitions of environment which naturally apply in these two approaches.

\section{Discussion and Summary}\label{sec:discussion}

We have applied the model of \cite{ludlow_mass-concentration-redshift_2016} to merger trees generated via the algorithm of \cite{parkinson_generating_2008} to predict the concentration of halos in those merger trees from their formation histories. We find that, with the original \cite{parkinson_generating_2008} algorithm this model results in insufficient scatter in concentration at fixed halo mass. Since the \cite{ludlow_mass-concentration-redshift_2016} model predicts the correct amount of scatter when applied to merger trees extracted from N-body simulations this suggests that the failure lies within the merger tree construction algorithm itself. We hypothesize that the missing ingredient is the effects of environment on halo formation times, and introduce a simple, empirical model for this effect to the \cite{parkinson_generating_2008} algorithm. With this modification our model correctly predicts the mean concentration as a function of halo mass, and comes closer to matching the measured scatter, while simultaneously matching the statistics of progenitor halo mass functions.

We also show that this simple model for the influence of environment on halo formation reproduces, at least qualitatively, the correlation between environment and formation epoch measured in N-body simulations. This brings some of the important effects of assembly bias into the extended Press-Schechter framework for merger tree construction.

\begin{figure}
\begin{tabular}{c}
\includegraphics[trim={0 0 0 0},clip,width=85mm]{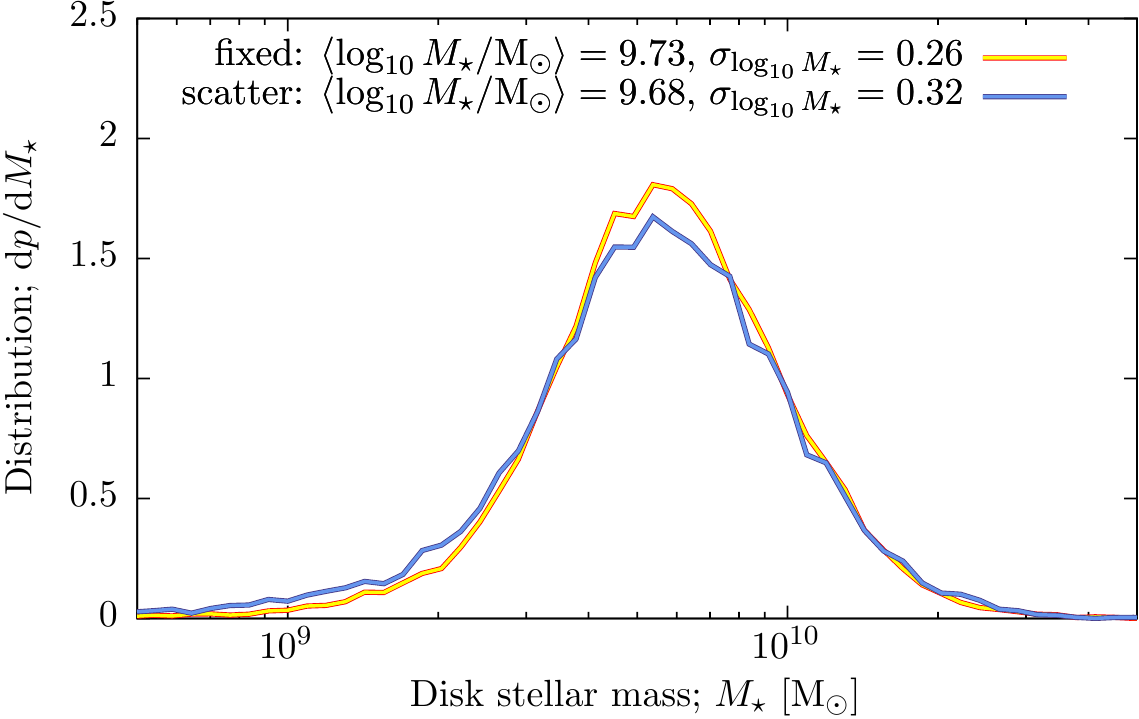} \\
\includegraphics[trim={0 0 0 0},clip,width=85mm]{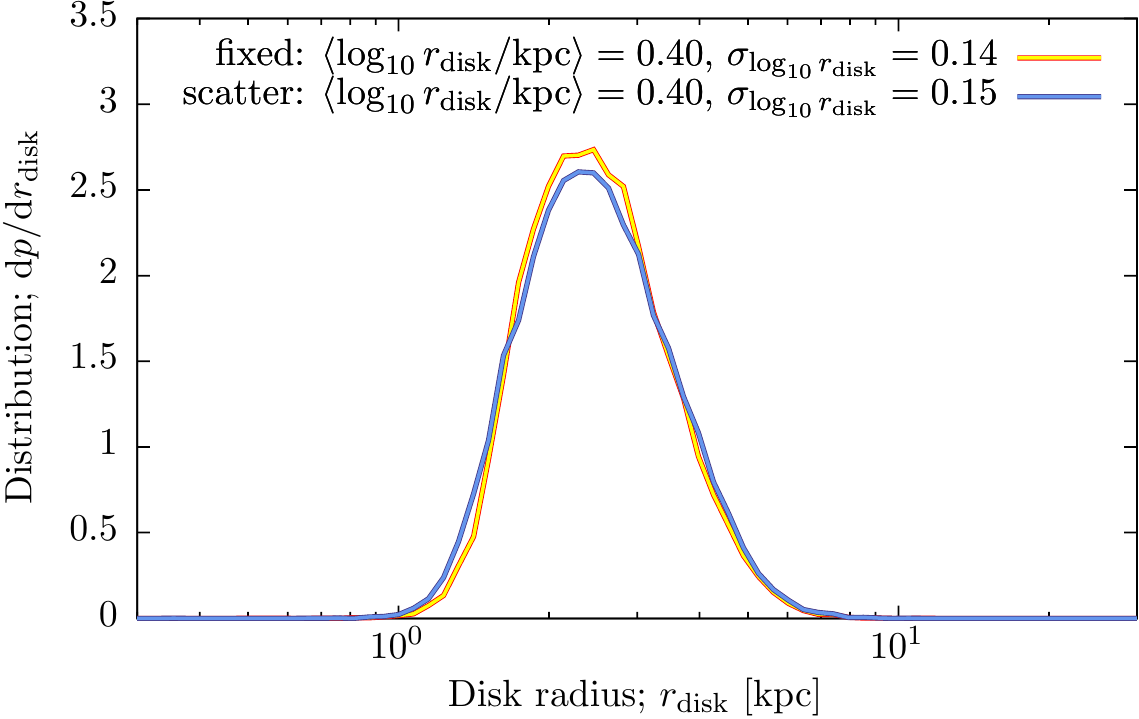} \\
\includegraphics[trim={0 0 0 0},clip,width=85mm]{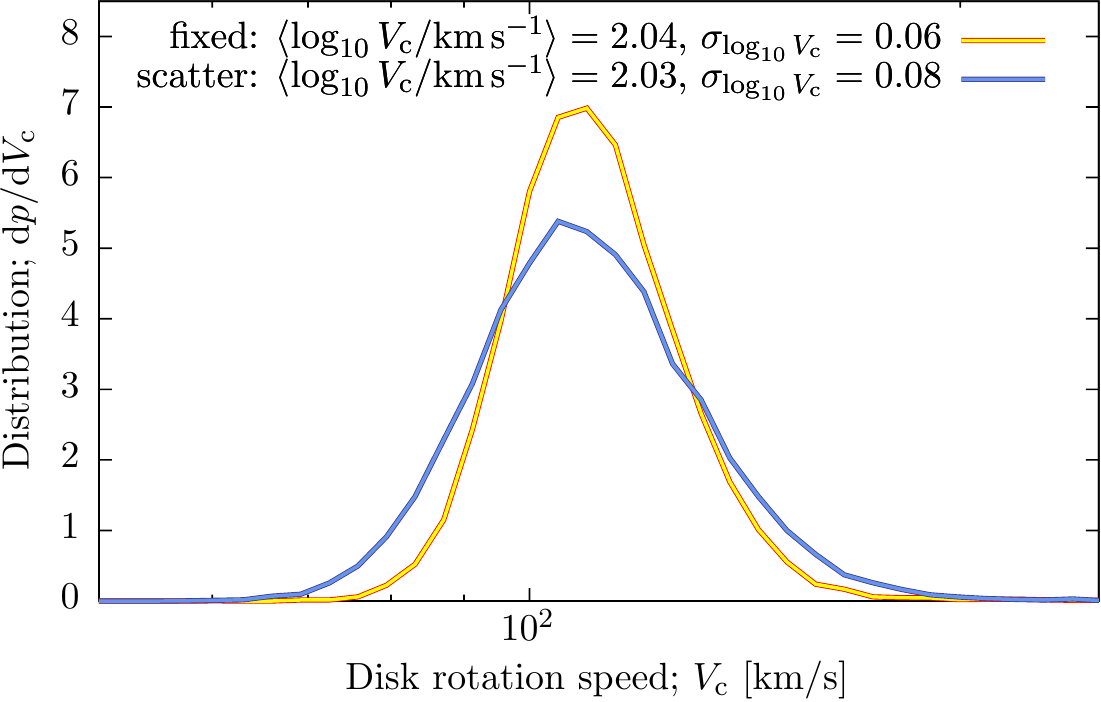}
\end{tabular}
\caption{The distribution of galactic disk stellar masses (upper), scale lengths (middle), and rotation speeds (at one scale length; lower) for disk-dominated, central galaxies occupying halos in the mass range 1--$3\times10^{12}\mathrm{M}_\odot$ at $z=0$. The yellow line shows the result obtained when the concentration of halos does not include any scatter (i.e. it is set to the mean concentration as a function of halo mass and redshift, specifically using the fitting formula given by \protect\cite{ludlow_mass-concentration-redshift_2016}). The blue line shows the result when scatter in concentration is included using the model of \protect\cite{ludlow_mass-concentration-redshift_2016} with parameter values as calibrated in this work. The mean and standard deviation of the distribution is shown for each case.}
\label{fig:distributions}
\end{figure}

This model has important consequences for semi-analytic models of galaxy formation built on such merger trees. The concentrations of halos are a key factor in determining the sizes of galaxies \citep{jiang_is_2018}, and determine key observable properties such as rotation curves. To illustrate the effects of this work we utilize the {\sc Galacticus} model \citep{benson_galacticus:_2012} to predict the distribution of disk stellar masses, scale lengths, and rotation speeds\footnote{In \textsc{Galacticus} galaxy sizes and rotation speeds are solved for assuming each disk is in rotationally-supported equilibrium in the combined gravitational potential of the dark matter halo and the baryonic content of the galaxy itself. The sizes and rotation speeds will therefore depend on the halo concentration through its effect on the dark matter density profile and potential. Stellar masses of galactic disks will also be affected by concentration since they depend on the sizes and rotation speeds of those disks (via the dependence of the algorithms for star formation and feedback on those quantities). See \protect\cite{benson_galacticus:_2012} for further details.} for central, disk-dominated galaxies occupying halos in the mass range 1--$3\times10^{12}\mathrm{M}_\odot$ at $z=0$. The resulting distributions are shown in Figure~\ref{fig:distributions}, which also shows the mean and standard deviation of each distribution. The yellow line shows the result obtained when the concentration of halos does not include any scatter (i.e. it is set to the mean concentration as a function of halo mass and redshift, specifically using the fitting formula given by \cite{ludlow_mass-concentration-redshift_2016}). The blue line shows the result when scatter in concentration is included using the model of \cite{ludlow_mass-concentration-redshift_2016} with parameter values as calibrated in this work. Incorporating scatter into halo concentrations has little effect on the mean of the distributions but does add to the scatter by a small but non-negligible increase. This is most readily apparent in the case of rotation speeds, in which the scatter is increased from 0.06 to 0.08~dex. Semi-analytic models which assign concentrations based on the mean concentration-mass relation will therefore miss a significant contribution to the scatter in these observables---as far as we are aware all other semi-analytic models either utilize a mean (or median) concentration-mass-redshift relation to assign concentrations to halos, or else do not make use of concentrations in deriving galaxy properties. Models which attempt to include this scatter by randomly drawing concentrations from the measured distribution of concentrations will, by construction, incorporate the scatter in observable quantities, but will miss any correlation with other galaxy properties that may be influenced by formation history.

\section*{Acknowledgements}

ADL acknowledges financial support from the Australian Research Council (project number FT160100250). We thank Nick Crook, whose initial investigation of applying the \cite{ludlow_mass-concentration-redshift_2016} model to PCH merger trees motivated this work. SMC acknowledges the support of the
Science and Technology Facilities Council [ST/P000541/1].

\bibliographystyle{mn2e}
\bibliography{concentrations}

\begin{thebibliography}{35}
\expandafter\ifx\csname natexlab\endcsname\relax\def\natexlab#1{#1}\fi

\bibitem[{Benson(2012)}]{benson_galacticus:_2012}
Benson A.~J., 2012, NewA, 17, 175

\bibitem[{Benson(2017{\natexlab{a}})}]{benson_constraining_2017}
---, 2017{\natexlab{a}}, MNRAS, 471, 2871

\bibitem[{Benson(2017{\natexlab{b}})}]{benson_mass_2017}
---, 2017{\natexlab{b}}, MNRAS, 467, 3454

\bibitem[{Benson {et~al.}(2012)Benson, Borgani, De~Lucia, Boylan-Kolchin, \&
  Monaco}]{benson_convergence_2012}
Benson A.~J., Borgani S., De~Lucia G., Boylan-Kolchin M., Monaco P., 2012,
  MNRAS, 419, 3590

\bibitem[{Benson {et~al.}(2013)Benson, Farahi, Cole, Moustakas, Jenkins,
  Lovell, Kennedy, Helly, {et~al.}}]{benson_dark_2013}
Benson A.~J., Farahi A., Cole S., Moustakas L.~A., Jenkins A., Lovell M.,
  Kennedy R., Helly J., {et~al.}, 2013, MNRAS, 428, 1774

\bibitem[{Bond {et~al.}(1991)Bond, Cole, Efstathiou, \&
  Kaiser}]{bond_excursion_1991}
Bond J.~R., Cole S., Efstathiou G., Kaiser N., 1991, ApJ, 379, 440

\bibitem[{Bond \& Myers(1996)}]{bond_peak-patch_1996}
Bond J.~R., Myers S.~T., 1996, ApJS, 103, 1

\bibitem[{Bullock {et~al.}(2001)Bullock, Kolatt, Sigad, Somerville, Kravtsov,
  Klypin, Primack, \& Dekel}]{bullock_profiles_2001}
Bullock J.~S., Kolatt T.~S., Sigad Y., Somerville R.~S., Kravtsov A.~V., Klypin
  A.~A., Primack J.~R., Dekel A., 2001, MNRAS, 321, 559

\bibitem[{Cole {et~al.}(2000)Cole, Lacey, Baugh, \&
  Frenk}]{cole_hierarchical_2000}
Cole S., Lacey C.~G., Baugh C.~M., Frenk C.~S., 2000, MNRAS, 319, 168

\bibitem[{Correa {et~al.}(2015)Correa, Wyithe, Schaye, \&
  Duffy}]{correa_accretion_2015}
Correa C.~A., Wyithe J. S.~B., Schaye J., Duffy A.~R., 2015, MNRAS, 452, 1217

\bibitem[{Davis {et~al.}(1985)Davis, Efstathiou, Frenk, \&
  White}]{davis_evolution_1985}
Davis M., Efstathiou G., Frenk C.~S., White S. D.~M., 1985, ApJ, 292, 371

\bibitem[{Einasto(1965)}]{einasto_construction_1965}
Einasto J., 1965, Trudy Astrofizicheskogo Instituta Alma-Ata, 5, 87

\bibitem[{Gao {et~al.}(2008)Gao, Navarro, Cole, Frenk, White, Springel,
  Jenkins, \& Neto}]{gao_redshift_2008}
Gao L., Navarro J.~F., Cole S., Frenk C.~S., White S. D.~M., Springel V.,
  Jenkins A., Neto A.~F., 2008, MNRAS, 387, 536

\bibitem[{Hellwing {et~al.}(2016)Hellwing, Frenk, Cautun, Bose, Helly, Jenkins,
  Sawala, \& Cytowski}]{hellwing_copernicus_2016}
Hellwing W.~A., Frenk C.~S., Cautun M., Bose S., Helly J., Jenkins A., Sawala
  T., Cytowski M., 2016, Mon Not R Astron Soc, 457, 3492

\bibitem[{Jiang {et~al.}(2018)Jiang, Dekel, Kneller, Lapiner, Ceverino,
  Primack, Faber, Macci\`o, {et~al.}}]{jiang_is_2018}
Jiang F., Dekel A., Kneller O., Lapiner S., Ceverino D., Primack J.~R., Faber
  S.~M., Macci\`o A.~V., {et~al.}, 2018, ArXiv e-prints, 1804, arXiv:1804.07306

\bibitem[{Klypin {et~al.}(2016)Klypin, Yepes, Gottl\"ober, Prada, \&
  He{\ss}}]{klypin_multidark_2016}
Klypin A., Yepes G., Gottl\"ober S., Prada F., He{\ss} S., 2016, MNRAS, 457,
  4340

\bibitem[{Lacey \& Cole(1993)}]{lacey_merger_1993}
Lacey C., Cole S., 1993, MNRAS, 262, 627

\bibitem[{Ludlow \& Angulo(2017)}]{ludlow_einasto_2017}
Ludlow A.~D., Angulo R.~E., 2017, MNRAS, 465, L84

\bibitem[{Ludlow {et~al.}(2016)Ludlow, Bose, Angulo, Wang, Hellwing, Navarro,
  Cole, \& Frenk}]{ludlow_mass-concentration-redshift_2016}
Ludlow A.~D., Bose S., Angulo R.~E., Wang L., Hellwing W.~A., Navarro J.~F.,
  Cole S., Frenk C.~S., 2016, MNRAS, 460, 1214

\bibitem[{Ludlow {et~al.}(2014)Ludlow, Navarro, Angulo, Boylan-Kolchin,
  Springel, Frenk, \& White}]{ludlow_mass-concentration-redshift_2014}
Ludlow A.~D., Navarro J.~F., Angulo R.~E., Boylan-Kolchin M., Springel V.,
  Frenk C., White S. D.~M., 2014, MNRAS, 441, 378

\bibitem[{Ludlow {et~al.}(2013)Ludlow, Navarro, Boylan-Kolchin, Bett, Angulo,
  Li, White, Frenk, {et~al.}}]{ludlow_mass_2013}
Ludlow A.~D., Navarro J.~F., Boylan-Kolchin M., Bett P.~E., Angulo R.~E., Li
  M., White S. D.~M., Frenk C., {et~al.}, 2013, MNRAS, 432, 1103

\bibitem[{Ludlow {et~al.}(2012)Ludlow, Navarro, Li, Angulo, Boylan-Kolchin, \&
  Bett}]{ludlow_dynamical_2012}
Ludlow A.~D., Navarro J.~F., Li M., Angulo R.~E., Boylan-Kolchin M., Bett
  P.~E., 2012, MNRAS, 427, 1322

\bibitem[{Maggiore \& Riotto(2010)}]{maggiore_halo_2010}
Maggiore M., Riotto A., 2010, ApJ, 711, 907

\bibitem[{Mo \& White(1996)}]{mo_analytic_1996}
Mo H.~J., White S. D.~M., 1996, MNRAS, 282, 347

\bibitem[{Navarro {et~al.}(1996)Navarro, Frenk, \&
  White}]{navarro_structure_1996}
Navarro J.~F., Frenk C.~S., White S. D.~M., 1996, ApJ, 462, 563

\bibitem[{Navarro {et~al.}(1997)Navarro, Frenk, \&
  White}]{navarro_universal_1997}
---, 1997, ApJ, 490, 493

\bibitem[{Navarro {et~al.}(2010)Navarro, Ludlow, Springel, Wang, Vogelsberger,
  White, Jenkins, Frenk, {et~al.}}]{navarro_diversity_2010}
Navarro J.~F., Ludlow A., Springel V., Wang J., Vogelsberger M., White S.
  D.~M., Jenkins A., Frenk C.~S., {et~al.}, 2010, MNRAS, 402, 21

\bibitem[{Parkinson {et~al.}(2008)Parkinson, Cole, \&
  Helly}]{parkinson_generating_2008}
Parkinson H., Cole S., Helly J., 2008, MNRAS, 383, 557

\bibitem[{Power {et~al.}(2003)Power, Navarro, Jenkins, Frenk, White, Springel,
  Stadel, \& Quinn}]{power_inner_2003}
Power C., Navarro J.~F., Jenkins A., Frenk C.~S., White S. D.~M., Springel V.,
  Stadel J., Quinn T., 2003, MNRAS, 338, 14

\bibitem[{Press \& Schechter(1974)}]{press_formation_1974}
Press W.~H., Schechter P., 1974, ApJ, 187, 425

\bibitem[{Roth \& Porciani(2011)}]{roth_testing_2011}
Roth N., Porciani C., 2011, MNRAS, 415, 829

\bibitem[{Sheth \& Tormen(2002)}]{sheth_excursion_2002}
Sheth R.~K., Tormen G., 2002, MNRAS, 329, 61

\bibitem[{Springel {et~al.}(2005)Springel, White, Jenkins, Frenk, Yoshida, Gao,
  Navarro, Thacker, {et~al.}}]{springel_simulations_2005}
Springel V., White S. D.~M., Jenkins A., Frenk C.~S., Yoshida N., Gao L.,
  Navarro J., Thacker R., {et~al.}, 2005, Nature, 435, 629

\bibitem[{Trenti {et~al.}(2010)Trenti, Smith, Hallman, Skillman, \&
  Shull}]{trenti_how_2010}
Trenti M., Smith B.~D., Hallman E.~J., Skillman S.~W., Shull J.~M., 2010, ApJ,
  711, 1198

\bibitem[{Zehavi {et~al.}(2018)Zehavi, Contreras, Padilla, Smith, Baugh, \&
  Norberg}]{zehavi_impact_2018}
Zehavi I., Contreras S., Padilla N., Smith N.~J., Baugh C.~M., Norberg P.,
  2018, ApJ, 853, 84

\end{thebibliography}

\appendix

\section{Optimization of the Merger Tree Construction Algorithm}\label{sec:cole2000Optimize}

We use the merger tree construction algorithm of \citeauthor{cole_hierarchical_2000}~(\citeyear{cole_hierarchical_2000}; see also \citealt{parkinson_generating_2008}). In that algorithm, small steps in ``time'' (actually in $w(t)\equiv\delta_\mathrm{c}(t)/D(t)$, where $\delta_\mathrm{c}$ is the critical threshold for spherical collapse and $D(t)$ is the linear growth factor, which serves the role of a time variable) are taken. These steps, $\delta w$, are required to be sufficiently small that the probability, $P = R \delta w$ (where $R$ is the rate of branching events per unit interval of $w$), of a branching event in any timestep is small, typically $P<\epsilon$ with $\epsilon=0.1$. Additionally, the timesteps must be small enough to ensure that subresolution accretion onto the halo is not too large during the timestep, and that the approximations of the \cite{cole_hierarchical_2000} merger rate are valid. In the limit of high branching rate (which occurs when the mass of the branch in question is much larger than the mass resolution), the step will be limited by branching.

We make a small modification to this algorithm. We remove the limit on the timestep due to branching rate, retaining the other limits on the timestep. Then, in each timestep, we find the interval to the next branching event by drawing a value at random from a negative exponential distribution with rate parameter $R$. If this interval exceeds the maximum allowed timestep, no branching occurs, and the timestep proceeds. If the interval is less than the maximum allowed timestep, branching occurs at that point. In the regime of high branching rates this approach allows for larger (by a factor $1/\epsilon$ on average) timesteps to be taken. Note that we do not have to concern ourselves in the subsequent timestep with the fact that no branching occurred in the previous timestep because of the memorylessness nature of the negative exponential distribution. That is, the distribution of branching intervals conditioned on the fact that no branching occurred in the previous timestep is just the same negative exponential distribution.

The increase in speed of the tree building will depend on both the algorithm implementation, and the tree resolution. In the case of the specific implementation of the \cite{cole_hierarchical_2000} algorithm, with a mass resolution equal to $5 \times 10^{-6}$ times the final halo mass, this optimization results in speed-ups by a factor of 1.2.

% Don't change these lines
\bsp	% typesetting comment
\label{lastpage}
\end{document}